\documentclass[12pt,english,aps,prb,reprint,amsmath,amssymb,onecolumn,notitlepage]{revtex4-2}
\usepackage[T1]{fontenc}
\usepackage[latin9]{inputenc}
\usepackage{color}
\usepackage{babel}
\usepackage{float}
\usepackage{amsmath}
\usepackage{amssymb}
\usepackage{graphicx}
\usepackage{microtype}
\usepackage[unicode=true,pdfusetitle,
 bookmarks=true,bookmarksnumbered=true,bookmarksopen=true,bookmarksopenlevel=1,
 breaklinks=false,pdfborder={0 0 0},pdfborderstyle={},backref=false,colorlinks=true]
 {hyperref}

\makeatletter
\usepackage{braket}
\usepackage{mathtools}
\usepackage{graphicx}

\usepackage{microtype}
\hypersetup{linkcolor = blue,
            urlcolor  = blue,
            citecolor = blue,
            anchorcolor = blue}

\makeatother

\begin{document}
\title{Dissipative Landau-Zener transition with decoherence rate}
\author{Le Tuan Anh Ho}
\email{chmhlta@nus.edu.sg}

\affiliation{Department of Chemistry, National University of Singapore, 3 Science Drive 3 Singapore 117543}
\author{Liviu Ungur}
\email{chmlu@nus.edu.sg}

\affiliation{Department of Chemistry, National University of Singapore, 3 Science Drive 3 Singapore 117543}
\author{Liviu F. Chibotaru}
\email{liviu.chibotaru@kuleuven.be }

\affiliation{Theory of Nanomaterials Group, Katholieke Universiteit Leuven, Celestijnenlaan 200F, B-3001 Leuven, Belgium}
\date{\today}
\begin{abstract}

An innovative microscopic model with a minimal number of parameters: tunneling splitting gap, external field sweeping velocity, and decoherence rate is used to describe dynamics of the dissipative Landau-Zener transition in the presence of the decoherence. In limiting cases, the derived equation of motion gives rise to the well-known Landau-Zener and Kayanuma formula. In a general case, the description demonstrates a non-monotonic flipping probability with respect to the sweeping velocity, which is also found in some other models. This non-monotony can be explained by considering the competition and timescale of the quantum tunneling, crossing period, and decoherence process. The simplicity and robustness of the theory offer a practical and novel description of the Landau-Zener transition. In addition, it promises an alternative method to the electron paramagnetic resonance in measuring the effective decoherence rate of relevant quantum systems. 

\end{abstract}
\maketitle
\global\long\def\hmt{\mathcal{H}}%
\global\long\def\vt#1{\overrightarrow{#1}}%

\global\long\def\chip{\chi'}%
\global\long\def\chipp{\chi''}%

\global\long\def\tn{\mathrm{tn}}%

\section{Introduction}

The Landau-Zener transition dynamics of a quantum system is a fundamental problem in physics and has various applications. These cover a wide range of fields, such as molecular magnets \citep{Gatteschi2006,Wernsdorfer1999}, quantum optics \citep{Bouwmeester1995}, chemical reactions \citep{hanggi1990reaction}, solid states artificial atoms \citep{petersson2010quantum,petta2010coherent}, or recently with nitrogen-vacancy center in diamond \citep{fuchs2011quantum,awschalom2007challenges}, single-molecule spin transistor \citep{Troiani2017}, and spin qubits \citep{Taran2019a,seidler2022conveyor,whaites2022adiabatic}.

The original Landau-Zener transition problem where an isolated particle of spin $1/2$ changes its states under a linear variation of the external magnetic field was exactly solved around one century ago \citep{Landau1932,Zener1932,Stueckelberg1932,Majorana1932}. Since then, many authors have considered more and more realistic versions of the problem, especially when this two-level system is in interaction with the surrounding environment and accordingly a dissipation of energy occurs \citep{Kayanuma1984,Kayanuma1984a,Kayanuma1985,Ao1991,Kayanuma1998,Pokrovsky2003,Saito2007,Nalbach2009,Nalbach2010,Kenmoe2013,Nalbach2013,Nalbach2014,Chen2020,Leuenberger2000a,Malla2022,Werther2019,Novelli2015,Huang2018,Javanbakht2015,Troiani2017,Taran2019a,Wubs2006,Pokrovsky2007,Sinitsyn2003,Wubs2005}. However, the primary question in these researches is essentially the same: how an environment with some specific characteristics influences the flipping probability between the two adiabatic/diabatic states during the sweeping through the avoided crossing point. In three seminal works several decades ago, Kayanuma proposed some simple stochastic models and was successful in calculating the transition probability for the Landau-Zener transition in the presence of either longitudinal or transversal Gaussian noise \citep{Kayanuma1984,Kayanuma1984a,Kayanuma1985}. The most interesting result from his works is probably an expression of the transition probability in the strong damping limit, hereinafter called Kayanuma formula, which significantly deviates from the Landau-Zener formula at small sweeping velocity. After the works of Kayanuma, more and more extended models covering a wide range of the environment noises, or different types of spin-bath coupling (transversal and/or longitudinal coupling) were also developed \citep{gefen1987zener,Ao1991,Wubs2006,Vitanov1996,Kayanuma1998,Pokrovsky2003,Pokrovsky2007}. Some works went even further by considering other types of baths from Ohmic/non-Ohmic bosonic bath to spin bath \citep{Saito2007,Sinitsyn2003} or fermionic bath \citep{Chen2020}. Some studied the effect of the measurement process as well \citep{Troiani2017,Novelli2015}. Additionally, other spin system with spin number larger than $1/2$ is also discussed \citep{Kenmoe2013}. A great deal of interesting results and techniques, both analytical and numerical ones, are provided in these theoretical works.

For isolated systems, the Landau-Zener formula clearly states that the tunneling splitting gap and the sweeping velocity of the external field are the only two quantities determining the flipping probability between two states of the quantum system \citep{Landau1932,Zener1932,Stueckelberg1932,Majorana1932,Ho2014}. The myriad of the succeeding researches on a two-level system coupling with the environment further clarifies that the coupling will cause decoherence of the phase between two states of the system. Due to this decoherence process, the flipping probability during the Landau-Zener transition may substantially change depending on many factors such as the type and magnitude of the coupling, or the correlation of the noise, or the spectral density of the environment, or some specific quantities of the used models. These approaches are fine, except that these decisive and somewhat environment-specific factors are typically difficult to measure/determine and accordingly applications of the findings from these approaches are most often limited. 

Considering the beauty in the simplicity of the original Landau-Zener problem and its solution, we  suggest taking a step back from recent approaches and asking a simpler question: given a decoherence rate of the quantum phase, a tunneling splitting gap, and a sweeping velocity of the driving field, what is the corresponding equation of motion and the flipping probability of the Landau-Zener transition? Besides its straightforwardness and simplicity, the main advantage of this approach is that it does not require detailed knowledge of the coupling between the spin system and the surrounding environment and/or measurement process. Hence, it leverages the applicability of this approach in reality. Recently, Troiani \emph{et al.} \citep{Troiani2017} and Taran \emph{et al. }\citep{Taran2019a} have successfully adopted this approach to elucidate the effect of the decoherence process on the Landau-Zener transition in real molecular spin systems. A major drawback in these works comes from the \emph{phenomenological} governing equation. Accordingly, a ``redundant'' average time parameter, which is interpreted as the finite time resolution of the experiment, was introduced into their model to make it work with the experimental data. 

The purpose of this paper is to introduce a neat microscopic description of the Landau-Zener transition dynamics of a spin system in the presence of the decoherence. A spin system in a weak interaction with the environment, which is often the case, is examined to derived the equation of motion for the transition.  We focus on finding a simple answer for the mentioned problem with only a minimal set of parameters: decoherence rate, tunneling splitting gap, and external field sweeping velocity. In the next section, we introduce the microscopic model used for the investigation of the Landau-Zener transition dynamics with a decoherence rate. The equation of motion of the transition is presented subsequently. Sec. \ref{sec:Solution-of-the} will then be dedicated to solving the governing equation of motion. Two limiting cases: coherent and incoherent Landau-Zener transition, and the general numerical solutions of the equation of motion of the Landau-Zener transition dynamics are examined. Some insights into the general behavior of the solution and its implications are also given in this section. We conclude the work with discussions and summary in the last section.

\section{Microscopic description of Landau-Zener transition with decoherence rate \label{sec:Microscopic-description-of}}

A multi-level spin system consisting of some doublets and singlets characterized by a spin number $S$ (or a total angular momentum number $J$) and a generic Hamiltonian in the diabatic (localized) basis is considered \citep{Garanin2011,Ho2017,Ho2022a,Ho2022b,Ho2022c}:

\begin{multline}
\hmt=\sum_{m^{\mathrm{th}}}\left(\varepsilon_{m}+\frac{W_{m}}{2}\right)\ket{m}\bra{m}+\left(\varepsilon_{m}-\frac{W_{m}}{2}\right)\ket{m'}\bra{m'}+\sum_{m^{\mathrm{th}}}\left(\frac{\Delta_{m}}{2}\ket{m}\bra{m'}+\frac{\Delta_{m}^{*}}{2}\ket{m'}\bra{m}\right)+\sum_{n^{\mathrm{th}}}\varepsilon_{n}\ket{n}\bra{n},
\end{multline}
 where $m$ ($n$) indicates the doublet $m^{\mathrm{th}}$ or singlet $n^{\mathrm{th}}$; $W_{m}$ is the energy bias induced by the magnetic field between two diabatic states $\ket{m}$ and $\ket{m'}$; and $\Delta_{m}$ is the tunneling splitting gap of the $m^{\mathrm{th}}$ doublet. 

As introduced in the introduction, we consider the case when the spin system $S$ weakly interacts with the surrounding environment. That is to say, the system dynamics can be described by the Redfield equation \citep{Blum1996,Garanin2011,Ho2017}. In the previous papers, using the semi-secular approximation \citep{Garanin2011,Ho2017} and the stationary limit for excited doublets/singlets \citep{Ho2022b,Ho2022c}, we have shown that the density matrix elements of the ground doublet is subject to the following equations: 

\begin{gather}
\frac{dX_{1}}{dt}=-\Gamma_{e}X_{1}-2\left(\Delta_{1r}\rho_{11'i}-\Delta_{1i}\rho_{11'r}\right),\label{eq:dX1}\\
\frac{d\rho_{11'r}}{dt}=-\gamma_{11'}\rho_{11'r}+W_{1}\rho_{11'i}-\frac{\Delta_{1i}}{2}X_{1},\label{eq:drho11'r}\\
\frac{d\rho_{11'i}}{dt}=-W_{1}\rho_{11'r}-\gamma_{11'}\rho_{11'i}+\frac{\Delta_{1r}}{2}X_{1},\label{eq:drho11'i}
\end{gather}
where $X_{1}=\rho_{11}-\rho_{1'1'}$ is the population difference between two diabatic states of the ground doublet. $\rho_{11'r}$, $\rho_{11'i}$, $\Delta_{1r}$, $\Delta_{1i}$ are respectively the real and imaginary component of $\rho_{11'}$ and $\Delta_{1}$. Meanwhile, $\Gamma_{e}$ is the relaxation rate of the ground doublet population difference when the ground doublet tunneling splitting gap $\Delta_{1}$ is zero, which can be seen clearly from Eq. \eqref{eq:dX1}; $\gamma_{11'}$ plays the role of the thermal decoherence rate (escape rate) of the ground doublet population \citep{Ho2022b}. Within the stationary limit for excited doublets/singlets, density matrix elements corresponding to the excited doublets/singlets are linear combinations of the ground doublet density matrix elements and hence share the same relaxation behavior \citep{Ho2022b}. Defining new variables $\rho_{r}\equiv\left(\Delta_{1i}\rho_{11'i}+\Delta_{1r}\rho_{11'r}\right)/\Delta_{1}$ and $\rho_{i}\equiv\left(\Delta_{1i}\rho_{11'r}-\Delta_{1r}\rho_{11'i}\right)/\Delta_{1}$ where $\Delta_{1}=\sqrt{\Delta_{1r}^{2}+\Delta_{1i}^{2}}$, Eqs. (\ref{eq:dX1}-\ref{eq:drho11'i}) becomes: 

\begin{align}
\frac{dX_{1}}{dt} & =-\Gamma_{e}X_{1}+2\Delta_{1}\rho_{i},\label{eq:dX1_new_var}\\
\frac{d\rho_{r}}{dt} & =-\gamma_{11'}\rho_{r}-W_{1}\rho_{i},\label{eq:drho_r}\\
\frac{d\rho_{i}}{dt} & =-\gamma_{11'}\rho_{i}+W_{1}\rho_{r}-\frac{\Delta_{1}}{2}X_{1},\label{eq:drho_i}
\end{align}

In previous papers \citep{Ho2022a,Ho2022b,Ho2022c}, we have discussed that $\Gamma_{e}$ is the effective relaxation rate via other canonical channels such as Orbach, Raman and direct process. Hence, we separate this relaxation effect from the solution of the above equations to study the change of population difference resulting from the Landau-Zener transition only. By substituting $X_{1}=xe^{-\Gamma_{e}t}$, $\rho_{i}=p_{i}e^{-\Gamma_{e}t}$, $\rho_{r}=p_{r}e^{-\Gamma_{e}t}$ into Eqs. (\ref{eq:dX1_new_var}-\ref{eq:drho_i}), we obtain the following \emph{key }system of equations: 
\begin{align}
\frac{dx}{dt} & =2\Delta_{1}p_{i},\label{eq:dx/dt}\\
\frac{dp_{r}}{dt} & =-\gamma_{d}p_{r}-W_{1}p_{i},\label{eq:dpr/dt}\\
\frac{dp_{i}}{dt} & =-\gamma_{d}p_{i}+W_{1}p_{r}-\frac{\Delta_{1}}{2}x,\label{eq:dpi/dt}
\end{align}
where $\gamma_{d}\equiv\gamma_{11'}-\Gamma_{e}$. Since the energy bias evolves linearly with time in the case of the Landau-Zener transition, we can substitute $W_{1}=vt$ into the above system of equations and solve it to find the variation of the population after crossing the avoided crossing point. It should be emphasized that this system of equations can be applied for any time-dependent function of $W_{1}\left(t\right)$ as well. 

Let take a look at the derived key system of equations, Eqs. (\ref{eq:dx/dt}-\ref{eq:dpi/dt}). Apparently, after removing the effect of the relaxation via canonical channels $\Gamma_{e}$, the latter quantity $\gamma_{d}$ behaves as the effective decoherence rate of the ground doublet density matrix elements. Thus, this system of equations can be considered as the governing equation for the Landau-Zener transition in the existence of a decoherence rate $\gamma_{d}$. Besides its simplicity, the novelty of this system of equations lies in two following facts: 1) it is derived from a microscopic model for a spin; and 2) it has only three parameters: tunneling splitting gap $\Delta_{1}$, sweeping velocity $v$, and the (effective) decoherence rate $\gamma_{d}$. 

To solve the above system of equations, we need to assign some initial conditions. Certainly, any initial conditions can be used. However, to be consistent with the typical Landau-Zener transition problem, we consider the ideal case where the initial conditions are $x\left(-\infty\right)=1$ and $p_{r}\left(-\infty\right)=p_{i}\left(-\infty\right)=0$. Additionally, it is supposed that the relaxation via excited doublets/singlets is negligible during the sweeping through the avoided crossing point. The effect of the relaxation caused by other relaxation processes can be accommodated into the final solution by multiplying by $e^{-\Gamma_{e}t}$.

\section{Landau-Zener transition with decoherence rate: limiting cases and numerical solution \label{sec:Solution-of-the}}

\subsection{Limiting cases}

\subsubsection{Coherent Landau-Zener transition: Landau-Zener formula}

The above system of equations of the Landau-Zener transition in the presence of a decoherence of rate $\gamma_{d}$, Eqs. (\ref{eq:dx/dt}-\ref{eq:dpi/dt}), can be transformed into a third-order differential equation of $x\left(t\right)$ as follows:
\begin{gather}
\frac{d^{3}x}{dt^{3}}+\left(2\gamma_{d}-\frac{1}{t}\right)\frac{d^{2}x}{dt^{2}}+\left(\gamma_{d}^{2}+\Delta_{1}^{2}+v^{2}t^{2}-\frac{\gamma_{d}}{t}\right)\frac{dx}{dt}+\left(\gamma_{d}-\frac{1}{t}\right)\Delta_{1}^{2}x=0.\label{eq:LZ with decoherence}
\end{gather}

The Landau-Zener transition takes place in an coherence manner in the original Landau-Zener problem \citep{Landau1932,Zener1932,Stueckelberg1932,Majorana1932}. That is to say, there is no decoherence and thus $\gamma_{d}=0$. The above equation then becomes: 
\begin{gather}
\frac{d^{3}x}{dt^{3}}-\frac{1}{t}\frac{d^{2}x}{dt^{2}}+\left(\Delta_{1}^{2}+v^{2}t^{2}\right)\frac{dx}{dt}-\frac{\Delta_{1}^{2}}{t}x=0.\label{eq:LZ eq}
\end{gather}

This is precisely the density matrix equation for the original Landau-Zener problem as shown in Eq. (24) of Ref. \citep{Kenmoe2013} or the Appendix of Ref. \citep{Vitanov1999} (Noted that both Ref. \citep{Kenmoe2013} and \citep{Vitanov1999} used a slightly different notations from ours where their tunneling splitting gap and sweeping velocity will correspond to $v/2$ and $\Delta_{1}/2$ in this work). Certainly, the value of $x\left(+\infty\right)$ obtained from Eq. \eqref{eq:LZ eq} should be the famous Landau-Zener formula $x\left(+\infty\right)=2\exp\left(-\frac{\pi\Delta_{1}^{2}}{2v}\right)-1$.

\subsubsection{Incoherent Landau-Zener transition: Kayanuma formula}

We consider another limiting case where the decoherence rate is so large that the Landau-Zener transition sweeps through the avoided crossing point entirely in an incoherent manner. That is to say, we can set the left-hand side of Eqs. (\ref{eq:dpr/dt}-\ref{eq:dpi/dt}) to zero, which results in: 
\begin{align*}
\frac{dx}{dt} & =-\frac{\Delta_{1}\gamma_{d}}{\gamma_{d}^{2}+W_{1}^{2}}x,\\
p_{r} & =\frac{\Delta_{1}}{2}\frac{W_{1}}{\gamma_{d}^{2}+W_{1}^{2}}x,\\
p_{i} & =-\frac{\Delta_{1}}{2}\frac{\gamma_{d}}{\gamma_{d}^{2}+W_{1}^{2}}x.
\end{align*}
Taking the integration of the first equation using the given initial conditions results in: 
\begin{equation}
x\left(t\right)=\exp\left[-\frac{\Delta_{1}}{v}\arctan\left(\frac{vt}{\gamma_{d}}\right)\right],
\end{equation}
which approaches the Kayanuma formula \citep{Kayanuma1984} at infinity:
\begin{equation}
x\left(+\infty\right)=\exp\left[-\pi\Delta_{1}/2v\right].
\end{equation}
It should be noted that this limiting case has been considered by Leuenberger and Loss in Ref. \citep{Leuenberger2000a}. As a more general equation, ours ought to reduce to the same equation and produce the same results as in Ref. \citep{Leuenberger2000a} for this incoherent Landau-Zener transition case. 

\subsection{Numerical solutions}

\begin{figure}[H]
\centering{}\includegraphics[width=0.7\textwidth]{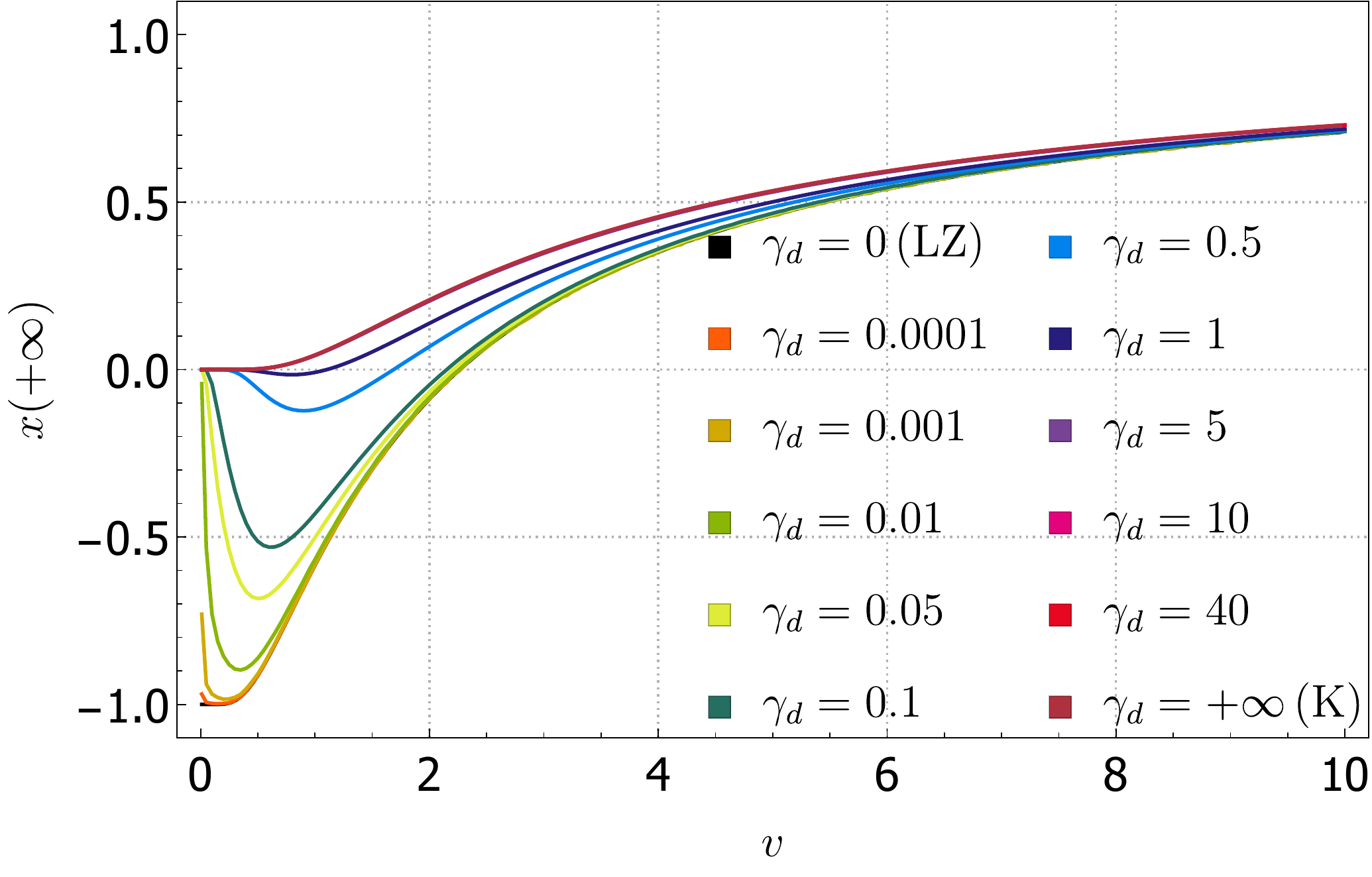}\caption{Dependence of the population difference $x\left(+\infty\right)$ between two diabatic states of the doublet on the sweeping rate $v$ with different values of $\gamma_{d}$ in $\Delta_{1}=1$ unit. The case $\gamma_{d}=0$ and $\gamma_{d}=+\infty$ respectively correspond to the Landau-Zener \citep{Landau1932,Zener1932,Majorana1932,Stueckelberg1932} and Kayanuma formula \citep{Kayanuma1984}. \label{fig:LZ_damping_velocity}}
\end{figure}

In order to gain more insight into the dynamics of the Landau-Zener transition with a decoherence rate, the main system of equations (\ref{eq:dx/dt}-\ref{eq:dpi/dt}) will be numerically examined. Its results then is compared with the Landau-Zener and Kayanuma formula. In Figure \ref{fig:LZ_damping_velocity}, we show the dependence of $x\left(+\infty\right)$ as a function of the sweeping velocity $v$ in a wide range of the decoherence rate in the tunneling splitting $\Delta_{1}=1$ unit. As can be seen from the figure, whereas the decoherence affects very little on the population difference $x\left(+\infty\right)$ at large sweeping velocity, a slight decoherence at low sweeping velocity will significantly modify the population difference $x\left(+\infty\right)$ from the well-known Landau-Zener formula. This can be qualitatively explained by considering the characteristic timescales of the coherence of the quantum tunneling of the population $\tau_{\mathrm{tunnel}}=1/\Delta_{1}$, of the decoherence $\tau_{\mathrm{decoherence}}=1/\gamma_{d}$, and of the crossing $\tau_{\mathrm{cross}}=\Delta_{1}/v$. In particular, a small sweeping velocity means that there is more time for the decoherence process to intervene in the coherent quantum tunneling of the population between two diabatic (localized) states during the crossing period, which then results in the loss of the quantum phase memory. This thus significantly diverges the flipping probability from the original Landau-Zener formula of coherence quantum tunneling. Surely the higher the decoherence rate, the stronger the decoherence is. Accordingly, the closer the flipping probability approaches the Kayanuma formula for the incoherence Landau-Zener transition. In the opposite case of fast sweeping velocity, the analogous explanation can be applied as well where the decoherence has less crossing time to make an impact on the coherent quantum tunneling. The numerical results in Fig. \ref{fig:LZ_damping_velocity} apparently confirm that the original Landau-Zener and Kayanuma formula indeed form the lower and upper limit for the Landau-Zener transition with decoherence. This totally makes sense considering that these two formulas respectively characterize for the case of fully coherent $\left(\gamma_{d}=0\right)$ and incoherent $\left(\gamma_{d}\rightarrow\infty\right)$ Landau-Zener transition. 

Interestingly, our numerical results clearly show the presence of a non-monotony in the transition probability with respect to the sweeping velocity. Although this non-monotony has been mentioned in some other works using different models/approaches \citep{Saito2007,Nalbach2009,Nalbach2010,Novelli2015,Chen2020,Nalbach2022}, it is intriguing to see that our simple model can reproduce this special feature. By considering three timescales $\tau_{\mathrm{tunnel}}$, $\tau_{\mathrm{decoherence}}$, and $\tau_{\mathrm{cross}}$ as previously, this feature can be qualitatively elucidated within our description.  In particular, a very small sweeping velocity $v$ and accordingly long $\tau_{\mathrm{cross}}$ will allow multiple quantum tunneling oscillations of the population between two diabatic states and the decoherence process characterized by $\tau_{\mathrm{decoherence}}$ has plenty of time to fully show its muscle. Consequently, the population difference $x\left(+\infty\right)$ will be close to the one given by Kayanuma formula. Increasing the sweeping velocity $v$ a little bit from zero/small value will then substantially reduce the relative duration of the crossing through the avoided crossing point. Accordingly, the decoherence process relatively has much less time to make an impact on the flipping probability. Roughly speaking, this decrease in the crossing duration is equivalent to reducing the decoherence, which then pushes the population difference $x\left(+\infty\right)$ toward one given by the coherent Landau-Zener transition, i.e. the Landau-Zener formula. This is clearly manifested on the left side of the Fig. \ref{fig:LZ_damping_velocity} where $x\left(+\infty\right)$ decreases as $v$ increases in the sweeping velocity domain close to zero. However, keeping increasing the sweeping velocity will then decrease $\tau_{\mathrm{cross}}$ closer to $\tau_{\mathrm{tunnel}}$. Taking the Landau-Zener formula of the flipping probability as an example, $P_{LZ}=1-\exp\left[-\frac{\pi\tau_{\mathrm{cross}}}{2\tau_{\mathrm{tunnel}}}\right]$, this $v$ increase results in a considerably fast decreasing of the flipping probability or equivalently a fast increase of $x\left(+\infty\right)$. Accordingly, it negates the effect of decreasing $x\left(+\infty\right)$ when increasing $v$ caused by the effective reduction of the decoherence rate as discussed above. In short, the opposite effects of increasing $v$ on the flipping probability due to 1) the effective reduction of the decoherence process, and 2) the decreasing of the effective number of population oscillations during the crossing interval are the reason behind the formation of a minimum in the domain where $\tau_{\mathrm{tunnel}}/\tau_{\mathrm{cross}}=\Delta_{1}^{2}/v\rightarrow\mathcal{O}\left(1\right)$ as can be seen in Fig. \ref{fig:LZ_damping_velocity}. 

\begin{figure}[H]
\centering{}\includegraphics[width=0.7\textwidth]{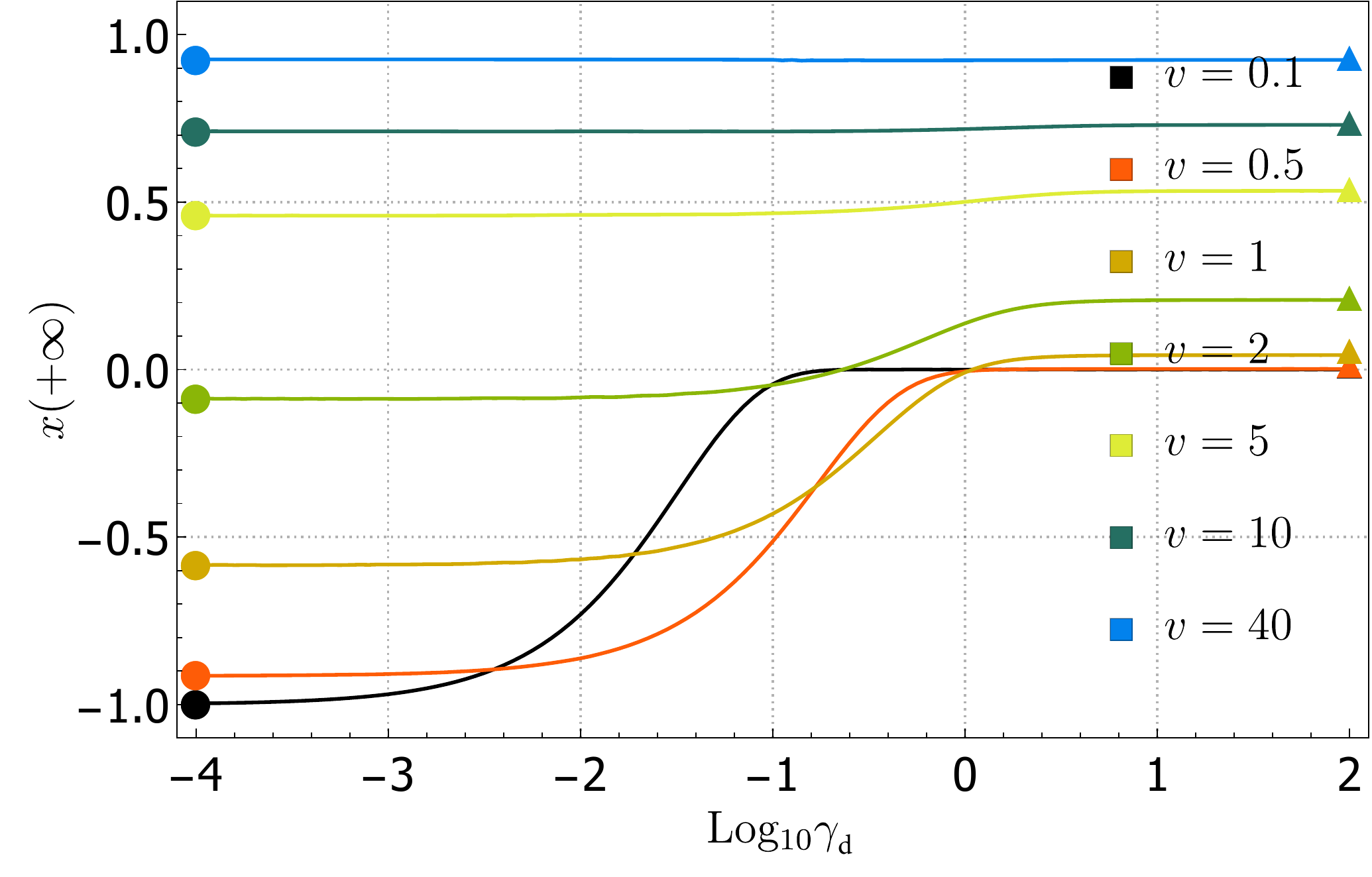}\caption{Dependence of the population difference $x\left(+\infty\right)$ between two diabatic states of the doublet on the decoherence rate $\gamma_{d}$ with different values of the sweeping rate $v$ in $\Delta_{1}=1$ unit. Circle and triangle marker respectively correspond to $x\left(+\infty\right)$ calculated from the Landau-Zener \citep{Landau1932,Zener1932,Majorana1932,Stueckelberg1932} and Kayanuma formula \citep{Kayanuma1984}. \label{fig:LZ_damping_decoherence rate}}
\end{figure}

In Fig. \ref{fig:LZ_damping_decoherence rate}, we investigate the dependence of the population difference $x\left(+\infty\right)$ on the decoherence rate $\gamma_{d}$ given the sweeping velocity. The most important observation from the figure is that $x\left(+\infty\right)$ is the most sensitive to the decoherence rate at low sweeping velocity $v$. This is easy to understand considering that the slower the sweeping velocity, the more time for the decoherence to exert its influence. This sensitivity thus provides a guidance for any measurement of the the decoherence rate using the Landau-Zener transition dynamics. Furthermore, Fig. \ref{fig:LZ_damping_decoherence rate} also reveals that the resolution of this decoherence rate measurement, if any, is less than about three orders of magnitude and the slower the sweeping velocity, the more accurate the measurement probably is. Accordingly, any fitting of the experimental data should take this sensitivity into account by giving more weight to samples at low sweeping velocity.

It is unsurprising from Fig. \ref{fig:LZ_damping_decoherence rate} that the values calculated by the Landau-Zener and Kayanuma formula set the lower and upper limit for $x\left(+\infty\right)$. As the decoherence rate is larger than the tunneling splitting $\Delta_{1}$ ($\log_{10}\gamma_{d}=0$ in Fig. \ref{fig:LZ_damping_decoherence rate}), it is safe to say that the system is subject to the Kayanuma formula and effectively behaves in an incoherent manner. On the other hand, the Landau-Zener formula is only justified as the decoherence rate is extremely smaller than the tunneling splitting and/or when the sweeping velocity is several times larger than the tunneling splitting. The latter comes from the fact that both the Landau-Zener and Kayanuma formula converges in this sweeping velocity domain. 

Lastly, although not as clear as in Fig. \ref{fig:LZ_damping_velocity}, we can still see from Fig. \ref{fig:LZ_damping_decoherence rate} the mentioned non-monotony in the low sweeping velocity domain and $\gamma_{d}$ is smaller than $\Delta_{1}$ where increasing the sweeping velocity will decrease the population difference $x\left(+\infty\right)$. In particular, this is manifested as the curve corresponding to $v=0.1$ (black line) is above the ones corresponding to $v=0.5$ (orange line) and 1 (dark yellow line) in the intermediate sweeping velocity domain ($\log_{10}\gamma_{d}\in\left[-2,-1\right]$), which is different from the collective behavior of other curves of higher sweeping velocities. 

\section{Discussions}

Up to now, we have assumed a linear varying energy bias $W_{1}$ between two diabatic states as in the well-known Landau-Zener problem. However, it should be kept in mind that the derived system of equations, Eqs. (\ref{eq:dx/dt}-\ref{eq:dpi/dt}), is relevant to any time-dependent form of the energy bias $W_{1}$, such as driving pulsed or periodic magnetic field. Additionally, the derived equation of motion also allows to calculate the flipping probability as a function of time besides the flipping probability at the (positive) infinity time.

Choosing the initial and final time at infinity is a convenient approximated choice in the Landau-Zener original problem considering that the Landau-Zener state flipping mainly occurs during the crossing period. However, due to the involvement of the relaxation in reality, the Landau-Zener transition flipping probability obtained from the theory may deviate from the ideal case. Hence, it is worth reminding that we should multiply the factor $e^{-\Gamma_{e}t}$, which covers the relaxation effect via the effective relaxation rate $\Gamma_{e}$, into the theoretical Landau-Zener flipping probability before interpreting the experiment date, especially if the relaxation deems non-negligible during the measurement period. 

In this work, we have mainly studied the Landau-Zener transition within the ground doublet. Some may raise a question about the Landau-Zener transition flipping probability within the excited doublets. Since the lifetime of the excited doublets is short either due to either very fast spontaneous emission to lower doublets at low temperature or fast population transfer at high temperature, the coherence between the states of these excited doublet is in fact rapidly phased out. Consequently, Landau-Zener transition within these excited doublets happens incoherently, i.e. $\gamma_{d}\rightarrow\infty$. The flipping probability, hence, should be subject to the Kayanuma formula in a large majority of cases. 

Our main equations, Eqs. (\ref{eq:dx/dt}-\ref{eq:dpi/dt}), are derived by considering a spin system in weak interaction with a thermal bath. Their similarity with Bloch equations and the role of the decoherence rate in these equation encourage us to make a pretty wild supposition that these equations can be used, at least to some phenomenological extent, for the Landau-Zener transition in the presence of the decoherence \emph{regardless} of the origin of the decoherence. That is to say, the decoherence may result from Gaussian noises, different types of baths (spin, Ohmic, non-Ohmic, etc.), different coupling strengths, or disturbance from the measurement process. In these cases, the parameter $\gamma_{d}$ needs to be conceived as the \emph{effective} decoherence rate. Some consistent hints on this supposition may be seen via the incoherent Landau-Zener transition with Kayanuma formula. In particular, despite that the nature of the model here and one considered by Kayanuma \citep{Kayanuma1984} are very different, they share the same final flipping probability in the incoherent Landau-Zener transition limit. Some other works with different models also produce the same flipping probability calculated by the Kayanuma formula for the incoherent Landau-Zener transition \citep{Ao1991,Pokrovsky2003}. These, to some extent, fortify our supposition.

In summary, we have derived a simple and intuitive framework to describe the dynamics of the Landau-Zener transition in the presence of the decoherence using only three parameters: the tunneling splitting gap, the sweeping velocity, and the decoherence rate. Our findings not only offer a handy and beautiful way to elucidate any deviation of the Landau-Zener transition experimental data from the well-known Landau-Zener or Kayanuma formula, but also add one more robust and effective method in determining the (effective) decoherence rate using Landau-Zener transition, besides the usual electron paramagnetic resonance (EPR) technique. The beauty in the simplicity of the theory and its corresponding equation of motion boosts up its applicability. Indeed, a fitting with experimental data can be easily made without unnecessary introduction of any redundant parameters. The theory is likely relevant for the Landau-Zener transition in a broad range of physical systems as well.

\begin{acknowledgments}
L. T. A. H. and L. U. acknowledge the financial support of the research projects R-143-000-A65-133, A-8000709-00-00, and A-8000017-00-00 of the National University of Singapore. Calculations were done on the ASPIRE-1 cluster (www.nscc.sg) under the projects 11001278 and 51000267. Computational resources of the HPC-NUS are gratefully acknowledged.
\end{acknowledgments}

\bibliographystyle{apsrev4-1}
\bibliography{references}

\end{document}